\documentclass[10pt,twocolumn]{article}

\usepackage[utf8]{inputenc}
\usepackage[T1]{fontenc}

\usepackage[small]{titlesec}

\usepackage{microtype}
\usepackage{newtxtext}
\usepackage{newtxmath}

\usepackage{cite}

\usepackage[left=2cm,right=2cm,top=2.5cm,bottom=2.5cm]{geometry}
\usepackage{listings}
\usepackage{xcolor} % for setting colors

\usepackage{enumitem}

\usepackage{graphicx}

\usepackage{hyperref}

\title{\textbf{stdgpu: Efficient STL-like Data Structures on the GPU}}

\author{Patrick Stotko\\
stotko@cs.uni-bonn.de\\
University of Bonn\\
}
\date{}

\definecolor{mainColor}{RGB}{30, 70, 230}
\definecolor{secondColor}{RGB}{140, 165, 250}

\definecolor{solarized@base03}{HTML}{002B36}
\definecolor{solarized@base02}{HTML}{073642}
\definecolor{solarized@base01}{HTML}{586e75}
\definecolor{solarized@base00}{HTML}{657b83}
\definecolor{solarized@base0}{HTML}{839496}
\definecolor{solarized@base1}{HTML}{93a1a1}
\definecolor{solarized@base2}{HTML}{EEE8D5}
\definecolor{solarized@base3}{HTML}{FDF6E3}
\definecolor{solarized@yellow}{HTML}{B58900}
\definecolor{solarized@orange}{HTML}{CB4B16}
\definecolor{solarized@red}{HTML}{DC322F}
\definecolor{solarized@magenta}{HTML}{D33682}
\definecolor{solarized@violet}{HTML}{6C71C4}
\definecolor{solarized@blue}{HTML}{268BD2}
\definecolor{solarized@cyan}{HTML}{2AA198}
\definecolor{solarized@green}{HTML}{859900}

\lstdefinestyle{stdgpuStyle}{
    language=C++,
    basicstyle=\scriptsize\ttfamily,
    tabsize=2,
    breaklines=true,
    escapeinside={@}{@},
    numberstyle=\tiny\ttfamily\color{solarized@base01},
    numbersep=6pt,
    otherkeywords={__host__, __device__, __global__},
    keywordstyle=\color{mainColor},
    stringstyle=\color{secondColor}\ttfamily,
    identifierstyle=\color{black},
    commentstyle=\color{solarized@base01},
    emphstyle=\color{solarized@red},
    frame=single,
    rulecolor=\color{solarized@base2},
    rulesepcolor=\color{solarized@base2},
    showstringspaces=false,
}

\hypersetup{colorlinks  = true,
            linkcolor   = black,
            citecolor   = black,
            filecolor   = black,
            urlcolor    = blue,
           }

\begin{document}

%\begin{titlepage}

\maketitle
%\tableofcontents

%\clearpage

%\end{titlepage}

% Remove page numbers on title page
%\thispagestyle{empty}

 \begin{abstract}
Tremendous advances in parallel computing and graphics hardware opened up several novel real-time GPU applications in the fields of computer vision, computer graphics as well as augmented reality (AR) and virtual reality (VR).
Although these applications built upon established open-source frameworks that provide highly optimized algorithms, they often come with custom self-written data structures to manage the underlying data.
In this work, we present \textit{stdgpu}, an open-source library which defines several generic GPU data structures for fast and reliable data management.
Rather than abandoning previous established frameworks, our library aims to extend them, therefore bridging the gap between CPU and GPU computing.
This way, it provides clean and familiar interfaces and integrates seamlessly into new as well as existing projects.
We hope to foster further developments towards unified CPU and GPU computing and welcome contributions from the community.
\end{abstract}

\section{Introduction}
\label{sec:introduction}

Efficiently solving complex problems often relies on carefully chosen data structures to process the data.
This allows to exploit structural properties of the problem and the data to achieve significant improvements regarding runtime performance or memory requirements.
In this context, several well-known data structures such as lists, sorted maps and sets, or hash-based structures were used.
The \textit{C++ Standard Libary} (\textit{C++ STL}~\cite{isostl}, also referred to as the \textit{Standard Template Libary}) provides a large variety of generic CPU data structures and algorithms for this purpose.

With the recent advances in graphics hardware, novel real-time applications in the field of computer vision and computer graphics~\cite{kinectfusion,kinectfusion2,niessner,infinitam,Dai:2017,Golodetz2018Collaborative,Reichl:2016}, augmented reality (AR) and virtual reality (VR)~\cite{stotko2019slamcast,mossel,Orts-Escolano:2016} were developed.
These sophisticated solutions make use of efficient data structures and algorithms to significantly improve the overall performance in comparison to the classic single-threaded applications.
However, the data structures do not naturally translate to the heavily parallelized GPU algorithms and are usually tailored to only fulfill the minimal application-specific requirements.
Consequently, modification operations such as insertion or removal functions and the missing concurrency between functions lead to race conditions or possible failures where e.g. the insertion of a value might fail in certain circumstances.
Previous solutions such as \textit{thrust}~\cite{bell2012thrust}, \textit{VexCL}~\cite{demidov2012vexcl}, \textit{ArrayFire}~\cite{malcolm2012arrayfire} and \textit{Boost.Compute}~\cite{szuppe2016boost} are focused on the fast and efficient implementation of various algorithms for contiguously stored data.
For this, they contain the implementation of a custom container data structure with interfaces and functionalities similar to the \texttt{std::vector} container.

\begin{figure}[t]
    \centering
    \includegraphics[width=\linewidth,height=\textheight,keepaspectratio]{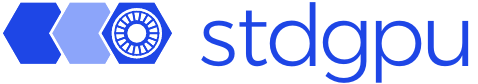}
    \caption{Logo of the \textit{stdgpu} library.}
    \label{fig:logo}
\end{figure}

In this report, we present the \textit{stdgpu} library which complements these approaches by providing several GPU data structures as the counterparts to the containers defined in the \textit{C++ STL}.
The logo of the library is depicted in Figure~\ref{fig:logo}.
It has been developed as part of the SLAMCast live telepresence system \cite{stotko2019slamcast} which performs real-time, large-scale 3D scene reconstruction from RGB-D camera images as well as real-time data streaming between a server and an arbitrary number of remote clients.
In comparison to other solutions, our data structures provide a large set of management functionality and guarantees to improve productivity while leveraging the high computational performance of modern GPUs.
Furthermore, our library is designed to extend and interoperate with existing frameworks using familiar interfaces to bridge the gap between CPU and GPU computing.
We released the \textit{stdgpu} library and made it available as open-source at \url{https://github.com/stotko/stdgpu}.

\section{Overview}
\label{sec:overview}

\begin{figure}[t]
    \centering
    \includegraphics[width=\linewidth,height=\textheight,keepaspectratio]{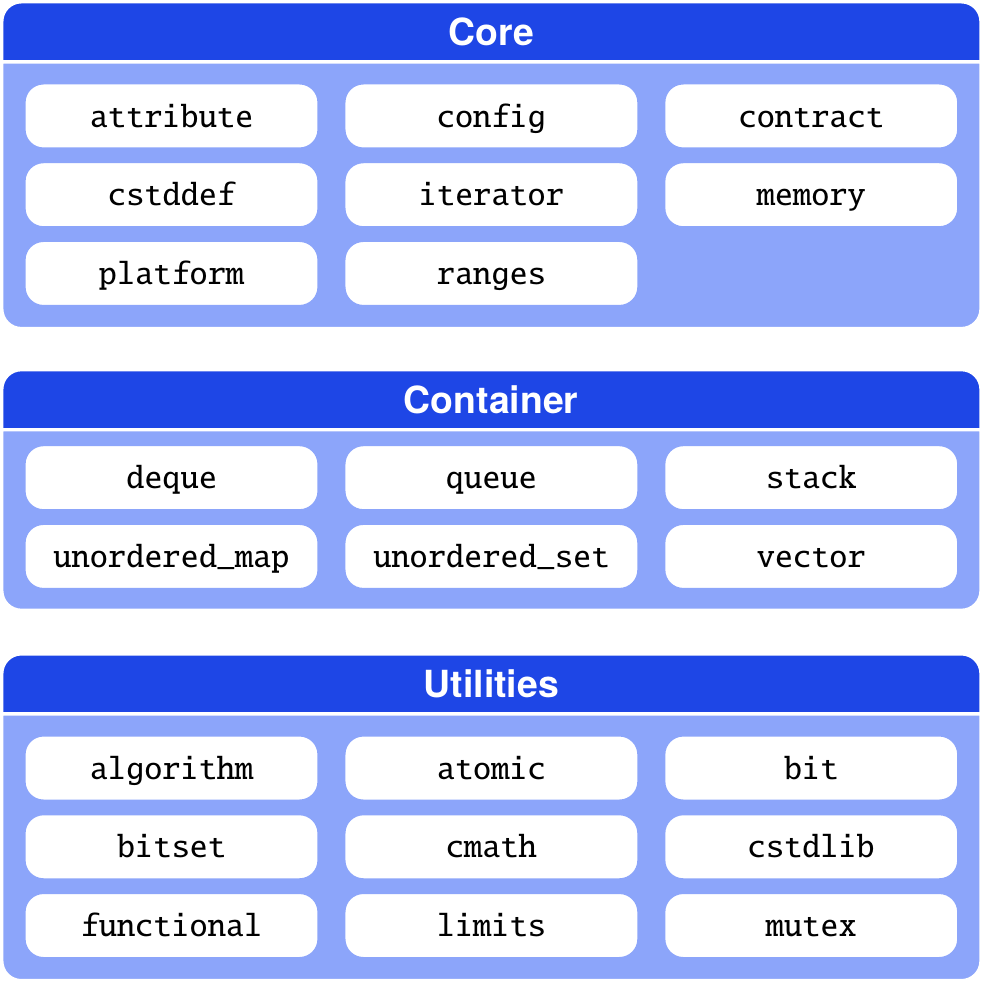}
    \caption{Overview of the features provided by the \textit{stdgpu} library divided into core, container, and utility components.}
    \label{fig:overview}
\end{figure}

Before the most important aspects of the \textit{stdgpu} library are described in detail, we first provide an overview of its features and structure as depicted in Figure~\ref{fig:overview}.
The main components of the library are:
\begin{itemize}[leftmargin=1em]\setlength\itemsep{0.0em}
    \item \textbf{Core}: A collection of core features including configuration and platform management, a simple contract interface as well as a robust memory and iterator concept.
    \item \textbf{Container}: A set of robust containers for GPU programming with an \textit{STL}-like design consisting of sequential and hash-based data structures.
    \item \textbf{Utilities}: A variety of utility functions supporting the container component and general GPU programming.
\end{itemize}

\section{Core}
\label{sec:core}

In order to build powerful libraries and applications, there are many design choices that focus on different aspects such as data and memory management.
The \textit{stdgpu} library also provides such a concept in its core component which is the foundation for the container and utility components.

\subsection{\texttt{platform}: Multi-Platform Support}
\label{sec:platform}

A crucial requirement for providing good compiler support as well as a high degree of usability and productivity is a solid platform concept that allows writing robust code for various compilers and contexts.
Here, we follow the common context terminology defined by the CUDA programming language and refer to code executed by the CPU as host code and code executed by the GPU as device code.
In order to provide cross-compiler support for functions that can be called by both the host and the device, we define the following function attribute:
\begin{lstlisting}[language=C++]
#if STDGPU_DEVICE_COMPILER == \
    STDGPU_DEVICE_COMPILER_NVCC
    #define STDGPU_HOST_DEVICE __host__ __device__
#else
    #define STDGPU_HOST_DEVICE
#endif
\end{lstlisting}
Here, \texttt{STDGPU\_DEVICE\_COMPILER} is the detected device compiler.
However, some algorithms require different code paths depending on the current context which would require defining two separate functions and calling the specialized host- or device-only version.
As the current context where the function is executed can also be detected, both versions can be unified to a single function:
\begin{lstlisting}[language=C++]
inline STDGPU_HOST_DEVICE int
some_function(const int x)
{
#if STDGPU_CODE == STDGPU_CODE_DEVICE
    // Device-specific implementation
#else
    // Host-specific implementation
#endif
}
\end{lstlisting}
This significantly simplifies writing unified CPU and GPU code.

\subsection{\texttt{cstddef}: Index Type Definition}
\label{sec:cstddef}

Another fundamental design choice for writing robust libraries and applications resides in the definition of the index type.
In particular, each unsigned and signed integer as well as 32-bit and 64-bit types have several advantages and disadvantages.
While the \textit{C++ STL} uses 64-bit unsigned integers, i.e. \texttt{std::size\_t}, for historical reasons, we follow another direction and use signed integers as our preferred index type:
\begin{lstlisting}[language=C++]
namespace stdgpu
{
using index32_t = std::int_least32_t;
using index64_t = std::ptrdiff_t;

#if STDGPU_USE_32_BIT_INDEX
    using index_t = index32_t;
#else
    using index_t = index64_t;
#endif
}
\end{lstlisting}
In comparison to unsigned integers which use modulo arithmetic for negative numbers, signed integers are less prone to common pitfalls and errors such as iterating over an empty array in reverse order.
Here, we also allow users to switch between sizes of 32 bits and 64 bits to optimize for performance in case very large indices do not occur.

\subsection{\texttt{contract}: Pre- and Post-Conditions}
\label{sec:contract}

Complex programs tend to be prone to subtle errors and bugs making it hard to detect such corner cases.
Many software development techniques try to mitigate this problem and improve the robustness and quality of the code.
We emulate contract programming techniques by defining pre- and post-condition checks using assertions.
This allows for specifying clear requirements and guarantees of functions which is demonstrated in the following example:
\begin{lstlisting}[language=C++]
float
safe_sqrt(const float x)
{
    STDGPU_EXPECTS(x >= 0.0f);

    float result = ...  // Compute the result

    STDGPU_ENSURES(result >= 0.0f);
    return result;
}
\end{lstlisting}
Here, the given function computes the square root of a real number which expects and ensures that both the input and output values are non-negative to obtain reasonable results.
Similar to the index definition, the checks can be disabled either manually by the user via the library configuration or automatically depending on the build type.

\subsection{\texttt{memory}: Robust Memory Management and Leak Detection}
\label{sec:memory}

For modern memory management, various containers and utility functions are defined in the \textit{C++ STL}.
\texttt{std::vector} and its equivalent versions in the \textit{thrust} library manage their allocated memory automatically making them significantly less prone to errors.
However, their application is limited to CPU-only code or a set of predefined algorithms.
In contrast, the goal of our container component is to support the usage in arbitrary GPU kernels and build more complex objects inheriting this property.
This bridges the gap between CPU and GPU programming and simplifies the development of simple, maintainable and fast code.
Therefore, we define a small set of functions for robust manual memory management which are used within the containers and provide strong guarantees regarding leak detection and operation safety.

\paragraph{Create and Destroy Arrays}
In order to allocate host or device memory, the following functions can be used:
\begin{lstlisting}[language=C++]
stdgpu::index_t n = 1000;
float x = 42.0f;

float* d_nums = createDeviceArray<float>(n, x);
float* h_nums = createHostArray<float>(n, x);
\end{lstlisting}
Here, two arrays of length $ n $ are created, one on the host site and one on the device site, and filled with the value $ x $.
In comparison to traditional allocations, these functions guarantee that the allocated memory is initialized with a given well-defined value.
At the end of the program, the allocated arrays must be deallocated to avoid memory leaks:
\begin{lstlisting}[language=C++]
destroyDeviceArray<float>(d_nums);
destroyHostArray<float>(h_nums);
\end{lstlisting}
In addition to the implicit filling, the arrays are registered to an internal leak detector which allows to catch double free errors and improve the safety of memory copies.

\paragraph{Copy Arrays between Host and Device}
Transferring data between the host and device is another essential operation and can be performed in the following way:
\begin{lstlisting}[language=C++]
stdgpu::index_t n = 1000;
float* h_nums = ... // Create array of length n
float* d_nums = ... // Create array of length n

// Fill host numbers with data
// ...

copyHost2DeviceArray<float>(h_nums, n, d_nums);

// Process the data on the device
// ...

copyDevice2HostArray<float>(d_nums, n, h_nums);
\end{lstlisting}
Here, the entire memory of \texttt{h\_nums} is copied to \texttt{d\_nums}.
After some processing, the data are copied back to the host.
A combination of allocation and copying is also provided:
\begin{lstlisting}[language=C++]
stdgpu::index_t n = 1000;
float* h_nums = ... // Create array of length n

// Fill host numbers with data
// ...

float* d_nums = copyCreateHost2DeviceArray<float>(h_nums, n);
\end{lstlisting}
The internal leak detector allows to check whether the participating arrays have been previously allocated and the memory range that should be copied is covered by the allocation bounds.
In order to support third party arrays, this check can be manually disabled by setting an additional parameter flag (default setting: on).

\subsection{\texttt{iterator}: Interoperability with Established Frameworks}
\label{sec:iterator}

Iterators are one of the core aspects of the \textit{C++ STL} and used to implement various algorithms such as counting, searching or sorting.
The \textit{thrust} library adopts this concept to define equivalent algorithms on the GPU.
Using our robust memory concept, we can request the size of allocated arrays to define a set of functions which allows to apply the iterator concept on plain arrays.
Thus, predefined algorithms can be used in a very robust and convenient way:
\begin{lstlisting}[language=C++]
float* nums = createDeviceArray<float>(1000);

// Fill the array with some data

thrust::sort(stdgpu::device_begin(nums),
             stdgpu::device_end(nums));
\end{lstlisting}
In this example, the start and end iterators of an array of numbers are determined and passed to a sorting algorithm.
For interoperability, the functions return dedicated pointer class types that encode the execution system, i.e. the host (CPU) or the device (GPU), which can then be inferred automatically inside the respective algorithm:
\begin{lstlisting}[language=C++]
namespace stdgpu
{
template <typename T>
device_ptr<T>
device_begin(T* d_array)
{
    return make_device(d_array);
}

template <typename T>
device_ptr<T>
device_end(T* d_array)
{
    return make_device(d_array + size(d_array));
}
}
\end{lstlisting}
This pointer class concept is also used to define iterator-based member functions for the container data structures.

\subsection{\texttt{ranges}: Towards Modern Interoperability}
\label{sec:ranges}

While iterators are a well-known concept and have been extensively used in the last decades, they only provide rather low-level semantics.
Many generic algorithms operate on a pair of iterators pointing to the start as well as the end of an array or container.
Modern approaches model these pairs as ranges which results in simpler and less error-prone code.
Due to the internal design of our container data structures (see Section~\ref{sec:container}), the notion of a start and end iterator pointing to the first as well as the past-the-end element may not always be well-defined.
However, the higher degree of flexibility of the range concept allows the construction of well-defined ranges for all containers.
The following example demonstrates how a selection operation, as required by e.g. 3D reconstruction and streaming techniques \cite{niessner,infinitam,stotko2019slamcast}, can be implemented:
\begin{lstlisting}[language=C++]
template <typename Functor>
void
select_blocks(const stdgpu::unordered_set<short3>& set,
              const Functor& selector,
              stdgpu::vector<short3>& set_selected)
{
    set_selected.clear();

    auto range = set.device_range();
    thrust::copy_if(range.begin(), range.end(),
                    stdgpu::back_inserter(set_selected),
                    selector);

    // Once range-based algorithms are provided:
    /*
    thrust::copy_if(set.device_range(),
                    stdgpu::back_inserter(set_selected),
                    selector);
    */
}
\end{lstlisting}
Here, the elements fulfilling the selection criterion are inserted into the \texttt{stdgpu::vector} container using a dedicated output iterator.
Using this design choice, we hope to encourage established frameworks such as the \textit{thrust} library to also follow this direction and provide a simple yet powerful range interface to generic algorithms.

\subsection{Host and Device Objects}
\label{sec:host_device_object}

Large projects tend to require more abstraction layers to keep the code maintainable and achieve a well-defined and concise design.
This includes defining data structures on top of plain arrays or containers in a possibly hierarchical fashion.
The containers defined in the \textit{stdgpu} library follow the object concept which can be considered as an extension of the memory interface:
\begin{lstlisting}[language=C++]
stdgpu::vector<int> d_vec = stdgpu::vector<int>::createDeviceObject(1000);

// Fill the vector with data and process them

stdgpu::vector<int>::destroyDeviceObject(d_vec);
\end{lstlisting}
In this example, an object of the \texttt{stdgpu::vector} container class with a capacity of $ 1000 $ elements is allocated on the device.
After some processing steps, the object is deallocated to avoid memory leaks.
In particular, objects are constructed and copied explicitly using the respective functions while the traditional copy constructor performs a shallow copy only.
Since GPU programs aim for maximum performance, this helps to clearly indicate intentional deep copies.

\section{Container}
\label{sec:container}

The heart of the \textit{stdgpu} library is its container component.
Built upon the core and utility components, it features the following design decisions:
\begin{itemize}[leftmargin=1em]\setlength\itemsep{0.0em}
    \item \textbf{Capacity}: Due to memory restrictions and performance reasons, data memory is preallocated using a fixed capacity specified on container creation. While manual resizing on the host can be used to mitigate this problem, we put our focus on the container operations.
    \item \textbf{Operations}: Inserting and removing data are the central interaction interface with the container and allow for efficient and robust data management in GPU kernels.
    \item \textbf{Concurrency}: In order to support a high degree of concurrency, the modification operations use a less strict locking system which permits failures of the current internal attempt but prevents potential deadlocks. These rare failures are resolved by further internal attempts to execute the operation. Using a strong invariant, read-only data lookup does not require any locking and is non-blocking.
    \item \textbf{Guarantees}: Despite thread-safe container modification, we also support concurrency across all defined operations. However, data insertion beyond the defined capacity is not supported and represents the only failure case.
\end{itemize}

\subsection{\texttt{unordered\_map} and \texttt{unordered\_set}: Hash-based Collections}
\label{sec:unordered_set_and_map}

Unordered maps and sets are associative containers and allow for very efficient modifications and lookups in constant time, i.e. $ \mathcal{O}(1) $.
Furthermore, they guarantee that entries with a specific key are only contained at most once which is crucial in many applications to maintain a consistent state.
Here, we implemented the containers \texttt{stdgpu::unordered\_map} and \texttt{stdgpu::unordered\_set} using a shared base class since the stored value type is the only major difference between them.
While \texttt{stdgpu::unordered\_map} stores a key-mapped pair, \texttt{stdgpu::unordered\_set} only considers the keys themselves.
For more details regarding the underlying data layout and implementation, we refer to our previous work~\cite{stotko2019slamcast}.

Both containers enable fast and reliable data management of sparsely distributed data.
Popular applications in the fields of computer vision, computer graphics as well as AR and VR include volumetric 3D reconstruction approaches~\cite{niessner,infinitam,Dai:2017,Golodetz2018Collaborative,Reichl:2016} as well as reconstruction-based live telepresence systems~\cite{mossel,stotko2019slamcast}.
These approaches exploit the sparsity of the 3D space to only store data around the observed 3D objects based on spatial hashing of a volumetric data representation using voxel blocks.
In order to use the \textit{stdgpu} library, this only requires the definition of an appropriate hash function~\cite{teschner2003optimized}:
\begin{lstlisting}[language=C++]
namespace stdgpu
{
template <>
struct hash<short3>
{
    inline STDGPU_HOST_DEVICE std::size_t
    operator()(const short3& key) const
    {
        return key.x * 73856093
             ^ key.y * 19349669
             ^ key.z * 83492791;
    }
};
}

// Spatial hash map for voxel block management
using block_map = stdgpu::unordered_map<short3, voxel*>;
\end{lstlisting}
Here, the discrete 3D block coordinates are multiplied with large prime numbers and then fused together with the bitwise XOR function to obtain a hash value.
The functionality of the respective application can now be implemented using the following set of retrieval and modification methods provided by both containers:
\begin{itemize}[leftmargin=1em]\setlength\itemsep{0.0em}
    \item \texttt{insert}: Function-based and iterator-based versions are defined. The former integrates a single value into the container and returns an iterator to the insertion position (or the end iterator if the value is already inserted) which is useful if subsequent operations depend on this result. On the other hand, the latter is a simple and convenient version to insert an array of values.
    \item \texttt{erase}: Removing values from the container can be performed similar to insertion in a function-based or iterator-based manner.
    \item \texttt{find/contains}: In addition to the aforementioned modification functions, values within the container can be accessed or queried. While \texttt{find} returns an iterator to the requested entry (or the end iterator if it was not found) that can be used by subsequent operations, the \texttt{contains} function only determines whether the value is present.
\end{itemize}
In the following, we show two example implementations for parts of the SLAMCast system's server architecture~\cite{stotko2019slamcast}.
The first one describes the update of a client's stream set which stores the set of blocks queued for streaming:
\begin{lstlisting}[language=C++]
class stream_set
{
public:
    void
    add_blocks(const short3* blocks,
               const stdgpu::index_t n)
    {
        set.insert(stdgpu::make_device(blocks),
                   stdgpu::make_device(blocks + n));
    }

    // Further functions

private:
    stdgpu::unordered_set<short3> set;
    // Further members
};
\end{lstlisting}
Since the update only requires the integration of the blocks, the iterator-based version is sufficient in this context.
The second example, however, is more complex and involves a dependency between two containers.
In particular, this function computes the set of updated blocks which is then passed to the stream set:
\begin{lstlisting}[language=C++]
__global__ void
compute_update_set(const short3* blocks,
                   const stdgpu::index_t n,
                   const block_map tsdf_block_map,
                   stdgpu::unordered_set<short3> mc_update_set)
{
    // Global thread index
    stdgpu::index_t i = blockIdx.x * blockDim.x + threadIdx.x;
    if (i >= n) return;

    short3 b_i = blocks[i];

    // Neighboring candidate blocks for the update
    short3 mc_blocks[8]
    = {
        short3(b_i.x - 0, b_i.y - 0, b_i.z - 0),
        short3(b_i.x - 1, b_i.y - 0, b_i.z - 0),
        short3(b_i.x - 0, b_i.y - 1, b_i.z - 0),
        short3(b_i.x - 0, b_i.y - 0, b_i.z - 1),
        short3(b_i.x - 1, b_i.y - 1, b_i.z - 0),
        short3(b_i.x - 1, b_i.y - 0, b_i.z - 1),
        short3(b_i.x - 0, b_i.y - 1, b_i.z - 1),
        short3(b_i.x - 1, b_i.y - 1, b_i.z - 1),
    };

    for (stdgpu::index_t j = 0; j < 8; ++j)
    {
        // Only consider existing neighbors
        if (tsdf_block_map.contains(mc_blocks[j]))
        {
            mc_update_set.insert(mc_blocks[j]);
        }
    }
}
\end{lstlisting}
In contrast to many other GPU hash map data structures, our container supports function-based insertion within kernels without sacrificing operation reliability and enables the efficient and robust implementation of such operations.

\subsection{\texttt{vector}: Resizable Contiguous Array}
\label{sec:vector}

Similar to previous frameworks, we provide a contiguous container which can be resized to handle a dynamically changing number of elements.
However, in contrast, our version can also be passed to custom kernels to enable conveniently and directly appending elements.
We define the following set of key operations for the \texttt{stdgpu::vector} data structure:
\begin{itemize}[leftmargin=1em]\setlength\itemsep{0.0em}
    \item \texttt{push\_back,emplace\_back}: Both of these thread-safe insertion functions can be used within kernels and append the given values to the end of the container. \texttt{emplace\_back} constructs the value on-the-fly from the given arguments.
    \item \texttt{pop\_back}: This function removes the value at the end of the container and returns it to the user.
    \item \texttt{operator[]}: Convenient access as for plain arrays is also provided.
\end{itemize}
Such a contiguous container can be used in situations where the exact size of the output cannot be exactly determined before, e.g. as in the case of the Marching Cubes algorithm~\cite{Lorensen:1987:MCH}.
This algorithm is used in volumetric 3D reconstruction~\cite{kinectfusion,kinectfusion2,niessner,infinitam,Dai:2017,Golodetz2018Collaborative} as well as the SLAMCast system~\cite{stotko2019slamcast} to extract triangular meshes from implicit 3D signed distance fields.
While the exact size of the regular grid is known here, the resulting number of triangles in the mesh after extraction depends on the shape and size of the encoded model.
Using the \texttt{stdgpu::vector} data structure, each thread in a kernel can compute the data-dependent number of triangles independently in parallel and append the generated data to the container.

\subsection{\texttt{deque}: Indexed Queue Data Structure}
\label{sec:deque}

This data structure defines the same functions as \texttt{stdgpu::vector} to allow random element access as well as the insertion and removal of elements from its end.
In addition, the same modification operations are provided to operate at its beginning which makes it more flexible:
\begin{itemize}[leftmargin=1em]\setlength\itemsep{0.0em}
    \item \texttt{push\_back,emplace\_back}: Elements are appended at the container's end.
    \item \texttt{push\_front,emplace\_front}: In these versions, elements are prepended at the beginning of the container.
    \item \texttt{pop\_back}: This function removes elements from the container's end.
    \item \texttt{pop\_front}: Here, elements are removed from the beginning of the container.
    \item \texttt{operator[]}: Convenient random element access is also provided.
\end{itemize}
Therefore, the container can be used as a stack data structure (last-in, first-out) as well as a traditional queue (first-in, first-out).

\section{Utilities}
\label{sec:utilities}

In this component, several low-level structures and functions are defined to simplify the implementation of kernels and the library's container component.

\subsection{\texttt{bitset}: Space-Efficient Indicator Array}
\label{sec:bitset}

The efficiency of the implementation of each container relies on the \texttt{stdgpu::bitset} data structure which models a fixed-sized array of bits.
In the internal implementation of the container data structures, each pre-allocated value is assigned an indicator to determine whether it holds user-provided data.
This avoids the definition of special placeholder values which might lead to undesired behavior in some situations.
Further use cases include e.g. the usage of such indicator arrays as high-resolution binary voxel grids to reduce the memory requirements of volumetric 3D reconstruction~\cite{Reichl:2016}.

\subsection{\texttt{mutex}: Towards Fine-Grained Synchronization}
\label{sec:mutex}

Another utility data structure that is needed to implement the container operations is a reliable locking mechanism to perform fine-grained synchronization of critical operations on the GPU.
Similar to \texttt{stdgpu::bitset}, we define a special array type for a sequence of mutexes.
Since guaranteed acquisition of a mutex using busy-waiting techniques may lead to deadlocks and implementing deadlock avoidance algorithms efficiently may require internal knowledge of the GPU architecture and driver, we only consider a simple solution without busy-waiting.
Thus, an attempt to acquire a lock may fail which needs to be considered by the respective operation as in the case of our container operations.
We believe that further advances in GPU architecture design will improve the reliability of this operation.

\subsection{Further Utilities}
\label{sec:further}

Besides the aforementioned utility structures, we also provide additional smaller functions and structures including wrapper classes to perform atomic operations on values (\texttt{atomic}), bit operations (\texttt{bit}) as well as basic hash functions (\texttt{functional}) and numeric properties (\texttt{limits}) on arithmetic types.

\section{Conclusion}
\label{sec:conclusion}

In this work, we presented \textit{stdgpu}, an open-source library which provides several generic GPU data structures.
Our data structures enable simple, fast and reliable data management to improve the productivity while leveraging the high computational performance of modern GPUs.
We designed our library to work with and extend established frameworks to bridge the gap between CPU and GPU computing.
However, several parts towards a complete \textit{C++ STL} on the GPU are still missing including the implementation of sorted containers such as \texttt{std::map} and \texttt{std::set}, an extensible backend system to support more platforms as well as some design limitations.
Nevertheless, we hope to foster further developments in this direction.
In future library versions, we will extend the support towards a more complete subset of the \textit{C++ STL} and welcome contributions and collaborations from the community.

\section*{Acknowledgements}
This work was supported by the DFG projects KL 1142/11-1 (DFG Research Unit FOR 2535 Anticipating Human Behavior) and KL 1142/9-2 (DFG Research Unit FOR 1505 Mapping on Demand).

\bibliographystyle{plain}
\bibliography{literature}

\end{document}